\documentstyle[aps,epsf,multicol]{revtex}

\def\breakon{\end{multicols}\widetext\vspace{-.2cm}
\noindent\rule{.48\linewidth}{.3mm}\rule{.3mm}{.3cm}\vspace{.0cm}}
 
\def\breakoff{\vspace{-.2cm}
\noindent
\rule{.52\linewidth}{.0mm}\rule[-.27cm]{.3mm}{.3cm}\rule{.48\linewidth}{.3mm}
\vspace{-.3cm}
\begin{multicols}{2}
\narrowtext} 

\begin{document}

\draft

\title{Gap Fluctuations in Inhomogeneous Superconductors}

\author{Julia S. Meyer$^{1,2}$ and B. D. Simons$^1$}

\address{$^1$ Cavendish Laboratory, Madingley Road, Cambridge CB3 0HE,
  UK\\ 
$^2$ Theoretische Physik III, Ruhr-Universit\"at Bochum, 44780 Bochum, Germany}

\date{\today}

\maketitle

\begin{abstract}
Spatial fluctuations of the effective pairing interaction between
electrons in a superconductor induce variations of the order parameter 
which in turn lead to significant changes in the density of states. In 
addition to an overall reduction of the quasi-particle energy gap,
theory suggests that mesoscopic fluctuations of the impurity potential 
induce localised tail states below the mean-field gap edge. Using a
field theoretic approach, we elucidate the nature of the states in the
`sub-gap' region. Specifically, we show that these states are
associated with replica symmetry broken instanton solutions of the 
mean-field equations.
\end{abstract}


\begin{multicols}{2}

\narrowtext

%%%%%%%%%%%%%%%%%%%%%%
\section{Introduction}
%%%%%%%%%%%%%%%%%%%%%%

The density of states (DoS) of a bulk s-wave superconductor exhibits a
quasi-particle energy gap and a singularity at the gap edge. While the
gap is robust with respect to the addition of non-magnetic impurities 
(Anderson theorem~\cite{Anderson}), the integrity of the gap is 
destroyed by the pair-breaking effect of time-reversal symmetry breaking 
perturbations~\cite{Maki63,deGennes64}, e.g.~magnetic 
impurities~\cite{AbGo61}, and parallel magnetic fields in thin 
films~\cite{Ov69}. (For a general review see, e.g., 
Refs.~\cite{Makirev,deGennes_book}). A second, and more direct
way of influencing the integrity of the gap is through the imposition
of quenched spatial fluctuations of the superconducting coupling 
constant~\cite{lo,IoLa81}. Physically, such inhomogeneities can be 
induced by dislocations, twin or grain boundaries, or compositional 
heterogeneity as found in superconducting alloys~\cite{Abrikosov}. In
these cases, fluctuations in the superconducting coupling constant 
are reflected in inhomogeneities of the order parameter which, in 
turn, induce a smearing of the quasi-particle energy gap. 

In all of the cases described above, the mechanism by which the 
quasi-particle energy gap is
suppressed follows a similar scheme and is described by the same 
phenomenology: at the mean-field level, each of the perturbations
above lead to a suppression of the quasi-particle gap edge. The 
form of this suppression is contained within the
Abrikosov-Gor'kov theory~\cite{AbGo61} which describes the rearrangement of
the ground state under the constraints imposed by the self-consistency
equation. While the physical mechanisms of gap suppression differ, the 
mean-field equations depend on a single dimensionless parameter
characterising the strength of the external perturbation (see below). 
Even at the mean-field level, it is found that if the 
perturbation is strong enough, the system is driven into a homogeneous 
gapless phase before the superconductivity is ultimately destroyed. 

After the pioneering work of Abrikosov and Gor'kov~\cite{AbGo61} (originally
formulated in the context of magnetic impurities in the dirty 
superconductor), it was realised that the integrity of the gapped 
phase is compromised even if the perturbation is 
weak~\cite{BaTr97,LaSi00,LaSi01}. 
Optimal fluctuations of the random impurity potentials can conspire to create 
quasi-particle states localised on the length scale of the coherence
length~$\xi$ at energies below the mean-field
gap, i.e.~in the presence of disorder, the system fragments into an
inhomogeneous phase in which `droplets' of localised sub-gap states 
are embedded in the superconducting background. A similar scenario arises 
in proximity coupled systems, i.e.~SN hybrid structures, where the 
superconductor induces a gap in the normal region~\cite{VaBr00,feigelmann00}.
In the most recent investigation, it was shown~\cite{LaSi01} that,
close to the mean-field energy gap edge~$E_{\rm gap}$, the nature of the
quasi-particle states (their structure and spectral density) are
universal, depending only on the relative separation from the edge, 
the dimensionality, and the single dimensionless parameter
characterising the strength of the perturbation. 

On this background, we consider below the influence of a spatially 
varying coupling constant~$g({\bf r})$ on the quasi-particle properties 
of a conventional disordered $s$-wave superconductor. As mentioned above,
such a program is not new: the same problem was investigated in an 
earlier work by Larkin and Ovchinnikov~\cite{lo}. However,
although our aims, and indeed many of our conclusions, are broadly 
similar to those of Ref.~\cite{lo}, this investigation is 
motivated by two considerations: firstly, the development of a 
quasi-classical approach within the framework of the non-linear 
$\sigma$-model (NL$\sigma$M) to explore the nature of the
quasi-particle states in the `sub-gap' region serves as a useful 
prototype for future studies of related `droplet phase' instabilities 
in other interacting theories such as that presented by the 
superconductor/insulator transition in the disordered interacting 
system~\cite{Ov73,SIT}. Secondly, in developing and applying the 
$\sigma$-model approach, we will find that the Lifshitz-type 
arguments~\cite{Lifshitz} invoked in Ref.~\cite{lo} to determine the
profile of the DoS in the sub-gap region are flawed. Indeed, the theory
developed below will expose a general scheme which establishes the
universality of `gap fluctuations' in the $d$-dimensional system 
(and which is in accord with the zero-dimensional results of 
Ref.~\cite{VaBr00}).

With this introduction, let us formulate the model superconducting
system which we will consider. Our starting point is the Gor'kov or 
Bogoliubov-de Gennes Hamiltonian 
\begin{eqnarray}
{\cal H}=\pmatrix{{\cal H}_0 & \Delta \cr \Delta & 
-{\cal H}_0^T}_{\sc ph}\,,
\label{ham}
\end{eqnarray}
where the subscript `${\sc ph}$' refers to the particle/hole space and
${\cal H}_0={\bf p}^2/2m-\epsilon_{\rm F}+V({\bf r})$.  Here, 
$\epsilon_{\rm F}$ denotes the Fermi energy and $V({\bf r})$ represents 
a quenched random impurity potential whose distribution is characterised 
by the mean scattering time~$\tau$. For simplicity, we consider the 
potential to be drawn from a Gaussian white noise distribution, 
$\langle V({\bf r})V({\bf r}')\rangle_V=(2\pi\nu_0\tau)^{-1}\delta({\bf r}
-{\bf r}')$, where $\nu_0$ is the average DoS of the normal system. 

The order parameter (chosen to be real) has to be determined 
self-consistently from the condition $g^{-1}({\bf r})\Delta({\bf r})
=\langle\psi_\uparrow({\bf r})\psi_\downarrow({\bf r})\rangle/\nu_0$. Following
Ref.~\cite{lo}, we will assume that the (inverse) 
coupling constant~$g^{-1}({\bf r})$ exhibits small fluctuations around 
an average value~$1/\bar g$. As with the random impurity potential, 
these fluctuations of the coupling constant $\delta(1/g)({\bf r})\equiv 
g_1({\bf r})$ are drawn from a Gaussian distribution with zero mean, 
and correlation 
\begin{eqnarray*}
\langle g_1({\bf r})g_1({\bf r}')\rangle_g=\phi(|{\bf r}-{\bf r}'|)
\end{eqnarray*}
(our specific choice of notation being borrowed from Ref.~\cite{lo}).
Here we will impose the condition $g_1({\bf r})\bar{g}\ll 1$ so that the
coupling constant remains positive everywhere. Furthermore, we will assume 
that the correlations are characterised by some correlation length~$r_c$ 
which determines the range of $\phi$. 

Qualitatively, the response of the ground state to inhomogeneities in 
the coupling constant depends sensitively on the range of the
correlations. If the correlation length~$r_c$ is much larger than the 
superconducting coherence length 
\begin{eqnarray*}
\xi=\left({D\over 2|\Delta|}\right)^{1/2}
\end{eqnarray*}
(where $D$ diffusion constant), then, the 
order parameter can smoothly adjust to the local value of 
$g^{-1}({\bf r})$. In this case $\Delta({\bf r})\sim g({\bf r})$, and 
the local DoS, $\nu({\bf r})$, is fixed by the local value of the order 
parameter~\cite{lo}. In the opposite limit, one expects the faster
fluctuations of the coupling constant to be rectified by the `proximity
effect' coupling of neighbouring superconducting regions. It is in this
limit that the system becomes sensitive to quasi-classical phase 
coherence processes. Therefore, to focus our discussion, in the
following, we limit consideration to the quasi-classical and dirty 
limits, where the energy scales are arranged in the hierarchy
\begin{eqnarray*}
\epsilon_{\rm F}\gg 1/\tau\gg \Delta\,.
\end{eqnarray*}

Before turning to the formalism, let us summarise the main conclusions 
of this investigation. Firstly, as anticipated by Ref.~\cite{lo}, it
is found that inhomogeneities of the coupling constant are reflected 
in inhomogeneities of the superconducting order parameter. Setting 
$\Delta({\bf r})=\bar{\Delta}+\Delta_1({\bf r})$, where $\bar{\Delta}$ 
represents the homogeneous component of the order parameter and 
$\Delta_1({\bf r})$ its spatial fluctuation, one finds that
\begin{eqnarray*}
\left\langle \Delta_1({\bf q}) \Delta_1(-{\bf q}) \right\rangle_g=
\bar{\Delta}^2\,\left\langle g_1({\bf q}) g_1(-{\bf q}) \right\rangle_g\,
f^2(|{\bf q}|).
\end{eqnarray*}
Here $f(|{\bf q}|)$ represents a dimensionless function which is determined
self-consistently (see below).

Now, if the correlation length of the coupling constant is short as
compared to the coherence length, $r_c\ll \xi$, one finds~\cite{lo} 
that the equation of motion for the average quasi-particle Green 
function obeys a local non-linear equation which has the canonical 
form obtained in the theory of gapless superconductivity by Abrikosov 
and Gor'kov~\cite{AbGo61}. Specifically, in the mean-field 
approximation, the BCS singularity is rounded off, and the DoS
exhibits a reduced quasi-particle energy gap~\cite{lo}
\begin{eqnarray*}
E_{\rm gap}=\bar\Delta(1-\eta^{2/3})^{3/2}\,,
\end{eqnarray*}
where
\begin{eqnarray*}
\eta\sim \phi(0)\left({r_c\over \xi\ln (\xi/r_c)^2}\right)^2\,
\end{eqnarray*}
is a dimensionless parameter characterising the strength of the
correlations of the superconducting order parameter. 
Note that, in the present case, the conditions $g_1({\bf r})\bar g\ll 1$ 
(i.e.~$\Delta_1({\bf r})\ll\bar\Delta$) and $r_c\ll\xi$ imply that 
$\eta\ll 1$. Thus, as one would expect, if the coupling constant remains
positive everywhere, the system remains in the gapped phase.

However, the conclusions of the mean-field analysis are modified
significantly by optimal fluctuations of the random impurity
potential. Such fluctuations, which appear as spatially inhomogeneous 
instanton configurations of the mean-field equation, show the gap
structure to be fragile: mesoscopic
fluctuations generate spatially localised states at energies below
the mean-field gap. Close to the mean-field gap edge~$E_{\rm gap}$, 
these states are confined to droplets of size~\cite{lo}
\begin{eqnarray*}
r_{\rm drop}(\epsilon)\sim\xi\left({E_{\rm gap}-\epsilon\over
\bar\Delta}\right)^{-1/4}\,,
\end{eqnarray*}
diverging as $\epsilon$ approaches $E_{\rm gap}$. Since
$r_{\rm drop}\gg \ell\gg \lambda_{\rm F}$, each of these regions is
characterised by an entire band of localised states confined to 
each droplet. To exponential accuracy the corresponding sub-gap DoS 
varies as 
\begin{eqnarray}
\nu(\epsilon)\sim\exp\!\left[-a_{\rm d}(\eta)\nu_0 DL^{d-2}
\left({\xi\over L}\right)^{\!d-2}\!\left(\frac{E_{\rm gap}\!-\!
\epsilon}{\bar\Delta}\right)^{\!\frac{6-d}4}\right]\!,
\label{dos}
\end{eqnarray}
where $a_{\rm d}(\eta)$ represents a known dimensionless function of 
the control parameter~$\eta$. This result, which is non-perturbative 
in the dimensionless conductance of the normal system $\nu_0
DL^{d-2}$, differs from that obtained in Ref.~\cite{lo} but, instead,
mirrors the scaling obtained in the study of sub-gap states in 
superconductors with magnetic impurities~\cite{LaSi00}. Later we will
argue that the energy scaling of the DoS is not accidental but is a
universal feature of the sub-gap states in the superconducting system
(c.f. Ref.~\cite{LaSi01}).

When $r_{\rm drop}> L$, the system enters a zero-dimensional regime. 
Here the expression for the DoS (\ref{dos}) applies with $d=0$. 
Reassuringly, this result is found to be in accord with the 
exact universal result predicted for zero-dimensional models which 
exhibit a square root singularity at the level of 
mean-field~\cite{VaBr00}. (The origin of this universality in the 
present scheme was discussed in Ref.~\cite{LaSi01}.)

%%%%%%%%%%%%%%%%%%%%%%%%%%%%%%%%%%%%%%%%%%%%%%%%%%%%%%%%%%%%%%%%%%%%%%%%%%%%%%%

The paper is organised as follows: in Sec.~\ref{sec2} we formulate the
quantum partition function of the disordered superconductor in the 
framework of a replica field theory. In the dirty limit, we show that 
the low-energy properties of the bulk superconducting system is contained
within a non-linear $\sigma$-model action. Taking into account the 
response of the quasi-particles to inhomogeneities in the BCS coupling 
constant, we obtain a renormalised action describing gap fluctuations.
Within a mean-field analysis we show that the quasi-particle energy gap
is suppressed but the integrity of the gap edge is preserved. Taking 
into account instanton configurations of the action, in section~\ref{sec3} 
we show that integrity of the gap edge is compromised. An analysis
of the fluctuations in the vicinity of the instanton configurations 
shows that optimal fluctuations of the impurity potential induce
sub-gap states localised on the length scale of the superconducting 
coherence length. Finally, in Sec.~\ref{conclusions} we conclude on
the universality of the results obtained here. 

%%%%%%%%%%%%%%%%%%%%%%
\section{Field theory of the inhomogeneous superconductor}
\label{sec2}
%%%%%%%%%%%%%%%%%%%%%%

The field theory approach to the study of weakly disordered 
systems~\cite{We79,EfLa80,Finkel83} has been discussed and reviewed 
extensively in the literature (see, e.g., Ref.~\cite{Efetov}). Its 
extension to the consideration of disordered superconducting systems 
follows straightforwardly~\cite{Oppermann,SIT,AST}. Therefore, here we 
will only
briefly summarise the main elements in the construction of the (replica) 
field theory of the disordered superconductor in the framework of the 
non-linear $\sigma$-model. Using this formulation, we will thereafter 
investigate the response of the superconducting system to
inhomogeneities in the BCS coupling constant.

%%%%%%%%%%%%%%%%%%%%%%%%%%%%%%%%%%%%%%
\subsection{Non-linear $\sigma$-model}
\label{subsec-NLSM}
%%%%%%%%%%%%%%%%%%%%%%%%%%%%%%%%%%%%%%

The starting point of the analysis is the coherent state path integral
for the replicated partition function~\cite{EfLa80,Finkel83}
\begin{eqnarray*}
{\cal Z}^N=\int \prod_{a=1}^N D\bar{\psi}^a D\psi^a e^{-\sum_{a=1}^N S[\psi^a]},
\end{eqnarray*} 
where $\psi^a$ represent Grassmann fields and, defining Matsubara 
frequencies for the Fermi system, $\epsilon_n=(2n+1)\pi/\beta$,
\begin{eqnarray*}
S[\psi^a]=\int d{\bf r} \sum_{n\sigma}\bar\psi_{n\sigma}^a (i\epsilon_n-
{\cal H}_0)\psi_{n\sigma}^a+S_I[\psi^a]
\end{eqnarray*}
with
\begin{eqnarray*}
S_I[\psi^a]=\frac{1}{\nu_0}\int_0^\beta d\tau \int d{\bf r}\; g({\bf r})
\;\bar\psi_\uparrow^a \bar\psi_\downarrow^a \psi_\downarrow^a 
\psi_\uparrow^a.
\end{eqnarray*}

To account for the symmetry properties of the system, it is convenient 
to enlarge the field space by introducing the four-component fields
${\Psi^a}^T=(\psi_\uparrow^a,\bar\psi_\downarrow^a,\psi_\downarrow^a,
-\bar\psi_\uparrow^a)/{\sqrt{2}}$. This incorporates a particle/hole as 
well as time-reversal (or charge conjugation) space. Decoupling the 
quartic BCS interaction with the introduction of the order 
parameter~$\Delta({\bf r})$ (chosen to be real), the total action
assumes the canonical form
\begin{eqnarray*}
S[\Psi^a]&=&\int d{\bf r} \sum_n \bar\Psi^a_n (i\epsilon_n\sigma^{\sc cc}_3-
{\cal H})\Psi^a_n\\
&& +\nu_0 \int_0^\beta d\tau \int d{\bf r}\, g^{-1}({\bf r}), 
\Delta^2({\bf r},\tau)
\end{eqnarray*}
where ${\cal H}$ represents the Gor'kov Hamiltonian defined 
above~(\ref{ham}). Here $\sigma^{\sc cc}_3$ is a Pauli matrix in the 
newly introduced charge conjugation space. 

To explore the low-energy properties of the superconducting system we
follow a standard route: as with normal disordered conductors, when 
subjected to an ensemble average over the random impurity
distribution, the functional integral over the Fermionic fields $\Psi$ 
can be traded for an integral involving matrix fields $Q$. Physically, 
the fields $Q$, which vary slowly on the scaling of the mean-free 
path $\ell$, describe the soft modes of density relaxation --- the 
diffusion modes. In the quasi-classical limit ($\epsilon_{\rm F}\gg 
1/\tau$), the action for $Q$ is dominated by the saddle-point field 
configuration. In the dirty limit ($1/\tau\gg\Delta$), the
saddle-point equation 
\begin{eqnarray*}
Q({\bf r})=\frac i{\pi\nu_0}\langle{\bf r}|\,\left(i\hat{\epsilon}
\sigma^{\sc ph}_3\otimes\sigma^{\sc cc}_3+\frac{\partial^2}{2m}+
\epsilon_{\rm F}+\frac i{2\tau}Q\right)^{-1}\,|{\bf r}\rangle\,,
\end{eqnarray*}
where $\sigma^{\sc ph}_3$ is a Pauli matrix in particle/hole space, admits
the solution $Q_0=\Lambda\otimes\sigma^{\sc ph}_3\otimes\sigma^{\sc cc}_3$, 
with $\Lambda_{nm}={\rm sgn}(\epsilon_n)\delta_{nm}$ and 
$[\hat{\epsilon}]_{nm}=\epsilon_n\delta_{nm}$. In the limit $\epsilon\to 0$,
the saddle point is not unique but spans an entire manifold parameterised
by the unitary transformations $Q=TQ_0T^{-1}$. Taking into account
slow spatial and temporal fluctuations of the fields, the low-energy
long-range properties of the weakly disordered superconducting system
are described by a non-linear $\sigma$-model 
action~\cite{Efetov,Oppermann,SIT,AST}
\breakon
\begin{eqnarray}
S[Q,\Delta]=\nu_0\int d{\bf r}\,\int_0^\beta \!d\tau \, g^{-1}({\bf r})
\,\Delta^2({\bf r},\tau)
+\frac{\pi\nu_0}{8}\int d{\bf r}\; {\rm tr}\,\left[D(\partial
Q)^2-4(\hat{\epsilon}\sigma_3^{\sc ph}\otimes\sigma_3^{\sc cc}+\Delta
\sigma_2^{\sc ph})Q\right]\,,
\label{S-gen}
\end{eqnarray}
\breakoff
\noindent
where the matrix field obeys the non-linear constraint $Q^2=\openone$. 
The Hermitian matrices~$Q$ obey the the symmetry relation 
$Q=\sigma^{\sc ph}_1\otimes\sigma^{\sc cc}_1 Q^T\sigma^{\sc ph}_1
\otimes\sigma^{\sc cc}_1$ reflecting the symmetry properties of the dyadic 
product $\Psi\otimes\bar\Psi$. Note that the fields carry replica ($a,b$) as 
well as Matsubara ($n,m$) indices, i.e.~$Q=Q^{ab}_{nm}$.

This completes the formulation of the disordered superconducting system
as a functional field integral involving the replicated NL$\sigma$M action.
Our interest here is in the thermodynamic DoS obtained as 
\begin{eqnarray*}
\nu(\epsilon) = -\pi^{-1} \int  {d{\bf r}\over V} \,\Im 
\left[G({\bf r},{\bf r};i\epsilon_n\!\to\!\epsilon^+)\right]\,.
\end{eqnarray*}
Making use of the analytic continuation $\ln {\cal Z}=\lim_{N\to0}{(
{\cal Z}^N-1)}/N$, it is straightforward to show that the impurity
averaged DoS can be obtained from the identity
\begin{eqnarray}
\left\langle\nu(\epsilon)\right\rangle={\nu_0\over 4}\int {d{\bf r}\over 
V} \lim_{N\to 0} \left\langle {\rm tr}\left[\Lambda\otimes 
\sigma_3^{\sc ph}\otimes\sigma_3^{\sc cc} Q P_{\epsilon\epsilon}\right]
\right\rangle_{Q,\Delta}\,,
\label{gendos}
\end{eqnarray}
where $\langle\cdots\rangle_{Q,\Delta}=\int DQ \int D\Delta \cdots 
e^{-S[Q,\Delta]}$ and $P_{\epsilon\epsilon}$ projects onto the
diagonal element $\epsilon\epsilon$.

%%%%%%%%%%%%%%%%%%%%%%%%%%%%%%%%%%%%%%
\subsection{Self-consistent fluctuations of the order parameter}
\label{subsec-sc}
%%%%%%%%%%%%%%%%%%%%%%%%%%%%%%%%%%%%%%

Following Ref.~\cite{lo}, our strategy will be to use the mean-field 
solution of the homogeneous problem as a platform to develop a perturbative
expansion of the self-consistent order parameter. Specifically, by 
finding the deviation $\delta\Delta({\bf r})=\Delta_1({\bf r})$ of the order
parameter from its mean value~$\bar\Delta$ to leading order in $g_1({\bf r})$, 
integrating out fast fluctuations of $Q$, and averaging over 
random configurations of $g_1({\bf r})$, we will obtain an effective 
action for the quasi-particle degrees of freedom of the superconducting system.
With this effective theory, we will again use a saddle-point analysis 
to explore the rearrangement of the ground state due to the inhomogeneous
coupling constant. At the mean-field level, the solution reveals a 
homogeneous renormalisation of the superconducting gap from its bare
value. On this background, we will find that the hard gap predicted 
by the mean-field theory is further softened by gap fluctuations 
which are accommodated in the effective field theory by inhomogeneous 
instanton configurations of the fields~$Q$. 

Following this program, we begin by subjecting the action to a 
saddle-point analysis. Varying the action~(\ref{S-gen}) with respect to 
$Q$ and $\Delta$, one obtains the coupled saddle-point equations
\begin{mathletters}
\begin{eqnarray}
&&D\partial(Q\partial Q)+\left[Q,\hat{\epsilon}\sigma_3^{\sc ph}\otimes
\sigma_3^{\sc cc}+\Delta\sigma_2^{\sc ph}\right]=0,\\
&&g^{-1}({\bf r})\Delta({\bf r})={\pi\over4\beta} {\rm tr}
\,\left[\sigma_2^{\sc ph}Q({\bf r})\right].
\end{eqnarray}
\label{spe}
\end{mathletters}
For a homogeneous coupling constant, these equations admit a homogeneous
solution for the order parameter and $Q$. However, for a general 
inhomogeneous configuration for $g^{-1}({\bf r})$, an exact solution 
is unavailable and an approximate scheme must be sought.

Applied to the saddle-point equations~(\ref{spe}) above, the Ansatz
\begin{eqnarray}
Q_{nm}=(\cos\hat\theta_n\sigma^{\sc ph}_3\otimes\sigma^{\sc cc}_3+
\sin\hat\theta_n\sigma^{\sc ph}_2)\delta_{nm}\,,
\label{param}
\end{eqnarray} 
where $\hat\theta_n={\rm diag}(\theta_n^1,\,\dots\,,\theta_n^N)$ 
is replica 
diagonal, leads to the coupled saddle-point equations
\begin{mathletters}
\begin{eqnarray}
&&D\partial^2\hat\theta_n({\bf r})-2\left(\epsilon_n
\sin\hat\theta_n({\bf r})-\Delta({\bf r})\cos\hat\theta_n({\bf r})\right)=
0\,,\\
&&g^{-1}({\bf r})\Delta({\bf r})={\pi\over\beta}\sum_n
\sin\hat\theta_n({\bf r})\,.
\label{SC} 
\end{eqnarray}
\end{mathletters}
In the following we will drop the Matsubara indices and only reinstate 
them when necessary. These equations can be identified as self-consistent 
Usadel equations~\cite{Usadel,lo73} for the average quasi-classical Green 
function in the presence of an inhomogeneous coupling constant. 
Specifically, the former represents the reorganisation of the ground state 
due to spatial inhomogeneities in the order parameter, while the second 
equation enforces the self-consistency condition imposed on the order 
parameter.

For a homogeneous coupling constant~$\bar{g}$, the mean-field equations 
are solved by a homogeneous replica symmetric Ansatz with 
\begin{eqnarray*}
\hat\theta_0=\arccos\left(\frac{\hat{\epsilon}}{\hat{E}}\right)\,, 
\end{eqnarray*}
where $\hat{E}^2=\hat{\epsilon}^2+\bar\Delta^2$ and $\bar\Delta=(\pi\bar
g/\beta)\sum_n\sin\theta_{0n}$. In this case, as expected, we simply 
recover the BCS solution~\cite{AST}. This result, being independent of 
disorder, is simply a manifestation of the Anderson
theorem~\cite{Anderson} on the 
level of the effective action --- in the non-interacting system, a weak 
non-magnetic impurity potential has no influence on the average DoS. 

To accommodate spatial fluctuations of the coupling constant and, with 
them, fluctuations of the order parameter one should, in principle, 
solve the non-linear set of equations self-consistently. Evidently, such 
a program is infeasible. Instead, following Ref.~\cite{lo}, taking the 
relative fluctuations of the coupling constant to be small, we look for 
a perturbative expansion of the mean-field equations. 
To develop the expansion, we 
set $\hat{\theta}=\hat{\theta}_0+\hat{\theta}_1({\bf r})$ and, accordingly, 
$\Delta=\bar\Delta+\Delta_1({\bf r})$, where both $\hat{\theta}_1$ and 
$\Delta_1$ are of order $g_1$. Expanding to first order in $g_1$, one 
obtains the coupled linear equations for $\hat{\theta}_1$ and $\Delta_1$,
\begin{mathletters}
\begin{eqnarray} 
&&D\partial^2\hat{\theta}_1-2\underbrace{(\hat\epsilon\cos\hat\theta_0+\bar\Delta
\sin\hat\theta_0)}_{\displaystyle \hat{E}}\hat\theta_1+2\Delta_1\cos\hat{\theta}_0
=0\,,
\label{SP1}\\
&&\Delta_1({\bf r})={\pi\bar{g}\over \beta}\sum_n\theta_{1n}({\bf r})
\cos\theta_{0n}-g_1({\bf r})\bar g\bar\Delta\,.
\label{SC1}  
\end{eqnarray} 
\end{mathletters}
Transforming Eq.~(\ref{SP1}) to the Fourier representation, one obtains
the solution
\begin{eqnarray*}
\hat{\theta}_1({\bf q})=2{\Delta_1({\bf q})\cos\hat\theta_0\over 
D{\bf q}^2+2\hat{E}} 
\end{eqnarray*}
which, when inserted into Eq.~(\ref{SC1}), yields 
\begin{eqnarray*} 
\Delta_1({\bf q})=-\frac{\bar\Delta}{\frac\pi{\bar\Delta\beta}\sum_n\!
\left(\sin\theta_{0n}-\frac{2\bar\Delta\cos^2\theta_{0n}}{D{\bf q}^2+2E_n}
\right)}g_1({\bf q})\,.
\end{eqnarray*} 
Finally, performing the Matsubara summation, one finds
\begin{eqnarray} 
\Delta_1({\bf q})\equiv -\bar\Delta  g_1({\bf q})f(|{\bf q}|)\,,
\label{delta1} 
\end{eqnarray} 
where, normalising the wavevector $\tilde{\bf q}=\xi{\bf q}$ by the 
coherence length, 
\begin{eqnarray*}
f(q)&=&2\tilde q^2\over \pi-(\tilde q^4-1)^{1/2}
\ln\left[{\tilde q^2-(\tilde q^4-1)^{1/2}\over 
\tilde q^2+(\tilde q^4-1)^{1/2}}\right]\nonumber \\
&=&\cases{1-{\pi \tilde q^2\over 4}+\dots\qquad &$\tilde q\ll 1$,\cr
{1\over\ln \tilde q^2}&$\tilde q\gg 1$.\cr}
\end{eqnarray*}
From this result, we obtain the response of the order parameter to
spatial variations of the BCS coupling constant. As expected, the low Fourier
components ($q\ll 1/\xi$) of the order parameter smoothly follow
spatial fluctuations of $g^{-1}({\bf r})$. Perhaps more surprising is the 
response of the order parameter to fast fluctuations. As one would
expect, these fluctuations are suppressed by the proximity effect.
However, as noted by Ref.~\cite{lo}, the attenuation scales only 
as $1/\ln \tilde q^2$.

\begin{figure}
\centerline{\epsfxsize=3.3in\epsfbox{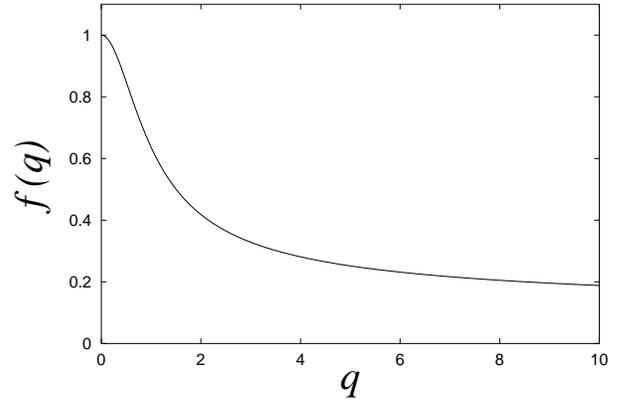}}
\caption{\label{fq} The function $f(q)$, governing the dependence of the 
order parameter on variations of the coupling constant.}
\end{figure}

Now, these fast fluctuations of the order parameter can have a
dramatic effect on the mean-field DoS and its fluctuation in the
vicinity of the mean-field gap edge. To assimilate the effect of these
fluctuations it is necessary to revisit the NL$\sigma$M
action taking into account the inhomogeneous order parameter. 

%%%%%%%%%%%%%%%%%%%%%%%%%%%%%%%%%%%%%%
\subsection{Mean-field solution}
\label{subsec-MF}
%%%%%%%%%%%%%%%%%%%%%%%%%%%%%%%%%%%%%%

Substituting the mean-field solution for the order parameter, $\bar\Delta$, 
together with its spatial fluctuation $\Delta_1({\bf r})$ into the 
NL$\sigma$M action, one obtains 
\begin{eqnarray*}
S[Q]=\frac{\pi\nu_0}8\int d{\bf r}\,{\rm tr}\left[D(\partial Q)^2-4\left(
\hat{\Sigma}+\Delta_1({\bf r})\sigma^{\sc ph}_2\right)Q\right]\,,
\end{eqnarray*}
where $\hat{\Sigma}=\hat\epsilon\sigma^{\sc ph}_3\otimes\sigma^{\sc cc}_3+
\bar\Delta\sigma^{\sc ph}_2$. 

Now, since $\int\!d{\bf r}\,\Delta_1({\bf r})=0$, contributions to the 
generating function arising from field configurations of $Q$ which are 
constant or slowly varying in space are largely insensitive to the 
fluctuations. Therefore, to assess the influence of the spatial 
inhomogeneity of the order parameter, we proceed by 
integrating out fast fluctuations of $Q$~\cite{Ef257/8}, where `fast' 
means varying on length scales shorter than the coherence length. 
To do so, the fast and slow degrees are separated by expanding $Q$ 
around the slowly varying $\bar Q$, i.e.  
\begin{eqnarray*} 
Q=T\bar QT^{-1}\,,\quad \bar Q=e^{-W_</2}\Lambda\otimes\sigma_3^{\sc ph}\otimes
\sigma_3^{\sc cc}\, e^{W_</2}\,,
\end{eqnarray*}
where $T=\exp[-W_>/2]$ with $\bar Q W_> +W_>\bar Q=0$. Integrating over $W_>$, 
one obtains $S_{\rm eff}=S_{\rm 0}+S_{\sc ag}$, where
\begin{eqnarray*}
S_{\rm 0}=\frac{\pi\nu_0}{8}\int d{\bf r}\; {\rm tr}
\,\left[D(\partial\bar Q)^2-4\hat{\Sigma} \bar Q\right]
\end{eqnarray*}
and
\breakon
\begin{eqnarray*}
S_{\sc ag}= \frac{\pi\nu_0}{8}\sum_{{\bf q},{\bf q}',{\bf q}'',{\bf q}'''}
{\rm tr}\,\left[{\Delta_1({\bf q})\Delta_1(-{\bf q}''')\,\hat{\Pi}_{{\bf q}',
-{\bf q}''}}[\bar{Q}({\bf q}+{\bf q}'),\sigma^{\sc ph}_2][\bar{Q}(-{\bf q}''
-{\bf q}'''),\sigma^{\sc ph}_2]\right]\,.
\end{eqnarray*}
Here $\Delta_1$ is determined by Eq.~(\ref{delta1}), and $\hat{\Pi}_{{\bf q},
-{\bf q}'}$ represents the diffusion propagator, i.e.~$\hat{\Pi}^{-1}_{{\bf q},
-{\bf q}'}=D{\bf q}^2\delta_{{\bf q},{\bf q}'}+\{\hat{\Sigma},\bar Q({\bf q}-
{\bf q}')\}$. Retaining only the diagonal part $\hat{\Pi}_{{\bf q},-{\bf q}}$ 
and averaging over fluctuations~$g_1$, which to a good approximation 
amounts to replacing $g_1({\bf q})g_1(-{\bf q}')$ by its average value 
$\langle g_1({\bf q}) g_1(-{\bf q}')\rangle_g=\phi(|{\bf q}|)\,
\delta_{{\bf q},{\bf q}'}$, one obtains 
\begin{eqnarray*}
S_{\sc ag}=\frac{\pi\nu_0\bar\Delta}{16}\int\!&& d{\bf r} \int d{\bf r}'
\;{\rm tr}\,\left[\hat{\eta}({\bf r}-{\bf r}',\hat\epsilon)\,[\bar Q({\bf r}),
\sigma^{\sc ph}_2][\bar Q({\bf r}'),\sigma^{\sc ph}_2]\right]
\end{eqnarray*}
\breakoff
\noindent 
with 
\begin{eqnarray*}
\hat{\eta}({\bf r},\hat\epsilon)=\frac2{\bar\Delta}\,\langle\Delta_1^2
\rangle({\bf r})\,\hat{\Pi}({\bf r}).
\end{eqnarray*}
Upon approaching the gap edge, $\hat{\Pi}$ becomes long-ranged, the 
relevant scale being $\ell_E=\sqrt{D/E}\gg\xi$. Thus, the spatial dependence 
of $\eta$ is governed by $\langle\Delta_1^2\rangle$, whose range is 
determined by Eq.~(\ref{delta1}) and the correlator $\langle g_1({\bf r})
g_1({\bf r}')\rangle_g=\phi(|{\bf r}-{\bf r}'|)$. If $\phi$ is short-ranged 
(on the scale of the coherence length), we can use the approximation 
\begin{eqnarray}
S_{\sc ag}=\frac{\pi\nu_0\bar\Delta}{16}\,\int d{\bf r}\;{\rm tr}\left(
\hat{\eta}(0,\hat\epsilon)\,[\bar Q({\bf r}),\sigma^{\sc ph}_2]^2\right)\,. 
\label{Snew}
\end{eqnarray}
Close to the gap, where the energy dependence of $\hat{\eta}$ is negligible, 
one has 
\begin{eqnarray*}
\hat{\eta}=\eta\simeq\frac1{\xi^2}\int \frac{d{\bf q}}{q^2}\,f^2(|{\bf q}|)
\phi(|{\bf q}|)\,.
\end{eqnarray*}
In this approximation, we will show below that the action recovers the 
mean-field equation obtained in Ref.~\cite{lo}.

To summarise, quenched inhomogeneities in the coupling constant induce 
spatial fluctuations of the order parameter which are accommodated by a
rearrangement of the quasi-particles in the superconducting condensate. 
In the disordered system, this rearrangement is governed by the same
Usadel equation that describe the proximity effect in hybrid SN systems.
Taking into account the inhomogeneities in the order parameter, one 
obtains an effective action for the disordered superconductor in which 
the bulk action for the non-disordered system is supplemented by an 
additional term~(\ref{Snew}) which, as we will see presently, leads to 
a suppression of the superconducting quasi-particle gap.

To explore the influence of the fluctuations on the quasi-particle gap 
structure let us again vary the action with respect to $\bar Q$. In doing
so, we obtain the modified saddle-point equation
\begin{eqnarray*}
D\partial(\bar Q\partial \bar Q)+\left[\bar Q,\hat{\Sigma}\right]-\frac12\bar\Delta \eta 
\left[\bar Q,\sigma_2^{\sc ph}\; \bar Q\; \sigma_2^{\sc ph} \right]=0\,.
\end{eqnarray*}
Adopting the parameterisation~(\ref{param}), this saddle-point equation can be 
rewritten as
\begin{eqnarray}
D\partial^2\hat\theta-2\left[\hat{\epsilon}\sin\hat\theta-\bar\Delta
\cos\hat\theta+\bar\Delta\eta\sin\hat\theta\cos\hat\theta\right]=0\,.
\label{SPn}
\end{eqnarray} 

The saddle-point equation~(\ref{SPn}) has a form which coincides with 
the Abrikosov-Gor'kov (AG) equation obtained in the theory of gapless 
superconductivity~\cite{AbGo61}. There, the parameter~$\eta$ has to be 
interpreted as the spin scattering rate $\eta=1/\tau_s\bar{\Delta}$ 
induced by magnetic impurities in the superconducting system. The
analysis of the AG equation shows that, for $\eta>1$ the system enters
a gapless phase while for $\eta<1$ the quasi-particle energy gap is 
suppressed but not destroyed. 

More precisely, taking the mean-field solution to be homogeneous in space,
i.e.~$\hat{\theta}({\bf r})\equiv\hat{\theta}_{\sc ag}$, the saddle-point
equation~(\ref{SPn}) takes the form
\begin{eqnarray*}
\hat\epsilon\sin\hat{\theta}_{\sc ag}-\bar\Delta\cos\hat{\theta}_{\sc ag}+
\bar\Delta\eta\sin\hat{\theta}_{\sc ag}\cos\hat{\theta}_{\sc ag} =0.
\end{eqnarray*}
Combined with the gap equation, the solution is obtained self-consistently 
from the equation
\begin{eqnarray*}
\frac{\hat{\epsilon}}{\bar\Delta}=\cot\hat{\theta}_{\sc ag}\left(1-\eta
\frac{1}{\sqrt{1+\cot^2\hat{\theta}_{\sc ag}}}\right)\,,
\end{eqnarray*} 
which coincides with Eq.~(18) of Ref.~\cite{lo}. Making use of the 
Eq.~(\ref{gendos}), one obtains the mean-field DoS
\begin{eqnarray*}
\nu(\epsilon)=\nu_0\,\Re\left[\cos\theta_{\sc ag}(i\epsilon_n\to\epsilon)
\right]\,,
\end{eqnarray*} 
which reveals a reduction of the energy gap according to
$E_{\rm gap}=\bar\Delta(1-\eta^{2/3})^{3/2}$. In particular, for
$\eta<1$, the mean-field theory still predicts a hard gap in the 
quasi-particle DoS, displaying square root behaviour as 
$\epsilon\to E_{\rm gap}$~\cite{lo}: 
\begin{eqnarray*}
\nu(\epsilon)\simeq \left\{\matrix{0&\enspace\epsilon<E_{\rm gap}\,,\cr\nu_0\,
\eta^{-2/3}(1\!-\!\eta^{2/3})^{-1/4}\sqrt{\frac23\,\frac{\epsilon-
E_{\rm gap}}{\bar\Delta}} &\enspace\epsilon>E_{\rm gap}\,.}\right.
\end{eqnarray*} 

As expected, quenched disorder in the coupling constant is reflected in
an overall suppression of the quasi-particle energy gap. Importantly, 
quasi-classical processes lead to a suppression of the gap that does not
simply follow the distribution of the order parameter. However, as
recognised by Larkin and Ovchinnikov~\cite{lo}, the square root 
singularity in the DoS predicted by the mean-field theory is untenable:
optimal fluctuations associated with the impurity potential~$V({\bf r})$
must give rise to sub-gap states which cause the gap to fluctuate. How
are such states accommodated by the statistical field theory? 

At first sight it is tempting to seek gap fluctuations within the 
perturbative fluctuations around the symmetric mean-field saddle-point
configuration $\theta_{\sc ag}$. However, when taken into account, it
is found that the integrity of the gap is maintained by the analytical
properties of the mean-field solution: perturbative fluctuations
influence only the profile of the DoS about the mean-field energy gap.
(For a discussion of this point in the context of the hybrid SN
system, see Ref.~\cite{AST}). Instead, to explore states below the
mean-field energy gap, it is necessary to revisit the saddle-point 
equation~(\ref{SPn}) and seek inhomogeneous `symmetry breaking' 
instanton field configurations. 

%%%%%%%%%%%%%%%%%%%%%%
\section{Inhomogeneous saddle points: Strategy}
\label{sec3}
%%%%%%%%%%%%%%%%%%%%%%

To develop a theory of sub-gap states in the present system we can draw
intuition both from the analysis of Larkin and Ovchinnikov~\cite{lo}
as well as a related study of gap fluctuations in the superconductor 
with magnetic impurities~\cite{LaSi00}. In Ref.~\cite{lo} sub-gap 
states were shown to be associated with inhomogeneous solutions of the 
Abrikosov-Gor'kov equation~(\ref{SPn}). In the framework of the field
theory of the non-interacting system, these inhomogeneous instanton or 
bounce solutions break supersymmetry at the level of the action. In the 
present case, we can therefore anticipate that the relevant bounce 
configurations
break the replica symmetry, providing an exponential suppression of the 
DoS below the mean-field edge. At the same time, we expect to identify
a zero-mode in the replica space which restores the global replica 
symmetry of the theory.

In the following, for simplicity, we will first restrict attention to the 
quasi one-dimensional system and describe later how the result can be 
generalised to higher dimensions.

%%%%%%%%%%%%%%%%%%%%%%%%%%%%%%%%%%%%%%
\subsection{Replica symmetry breaking}
\label{subsec-rsb}
%%%%%%%%%%%%%%%%%%%%%%%%%%%%%%%%%%%%%%

In order to investigate inhomogeneous solutions, we first have to
understand the structure of the homogeneous solution in more detail. 
The Fermionic
integration contour covers the interval $\theta\in[0,\pi]$. As the 
mean-field DoS vanishes below the gap edge (i.e.~$\nu(\epsilon)=
\nu_0\Re[\cos\theta]=0$), one
knows that the mean-field solution of the Abrikosov-Gor'kov equation
satisfies the condition $\theta_{\sc ag}(\epsilon<E_{\rm gap})=\pi/2+
i\phi_{\sc ag}(\epsilon)$, where $\phi_{\sc ag}(\epsilon)$ is real. 

\begin{figure}
  \centerline{\epsfxsize=3.3in\epsfbox{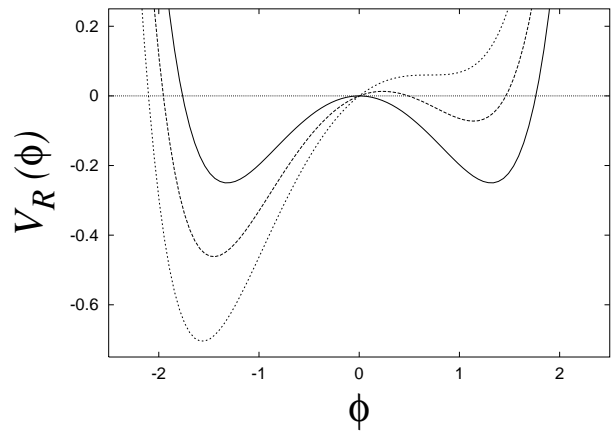}}
\caption{\label{potential} Real potential as a function of $\phi$
along the line $\theta=\pi/2+i\phi$ for $\epsilon=0,E_{\rm gap}/2,
E_{\rm gap}$. At $\epsilon=E_{\rm gap}$ the maximum and the minimum merge.}
\end{figure}

Now instead of analysing the saddle-point equation, it is more convenient 
to study its first integral 
\begin{eqnarray*} 
\xi^2(\partial\theta)^2+V(\theta)={\rm const.}\,, 
\end{eqnarray*}
where, switching from Matsubara to real energies, the effective potential 
is given by
\begin{eqnarray*}
V(\theta)=-2i\frac{\epsilon}{\bar\Delta}\cos\theta+2\sin\theta+\frac\eta2
\cos2\theta\,.
\end{eqnarray*}
Along the line $\theta=\pi/2+i\phi$ the potential $V_R(\phi)=-V(\pi/2+i
\phi)$ is {\em real} with a functional dependence on $\phi$ shown in 
Fig.~\ref{potential} for different energies. The homogeneous
saddle point sits at the maximum of this potential. At $\epsilon=0$,
it belongs to the contour ($\theta_{\sc ag}=\pi/2$) and the potential
is symmetric around $\phi=0$. By increasing the energy, the saddle
point moves away from the real axis along the line $\theta_{\sc ag}=\pi/2
+i\phi_{\sc ag}$ until the energy reaches $E_{\rm gap}$. On this line
the mean-field DoS vanishes. At energies $\epsilon>E_{\rm gap}$ the real 
part of $\theta_{\sc ag}$ starts to deviate from $\pi/2$ and the DoS 
becomes finite. Thus, even at the mean-field level the integration
contour must be smoothly deformed to reach the saddle-point solution
(for a discussion see, e.g., Ref.~\cite{AST}). Following the behaviour
of the potential, one notices that one minimum deepens while the other
becomes more and more shallow until merging with the maximum at
$\epsilon=E_{\rm gap}$.

Now, in addition to the homogeneous Abrikosov-Gor'kov solution, the potential 
above admits for a bounce solution \mbox{$\phi_{\sc ag}\to\phi_{\rm max}(
>\phi_{\sc ag})\to\phi_{\sc ag}$}, where $V_R(\phi_{\rm max})=
V_R(\phi_{\sc ag})$. In principle, as is
clear from Fig.~\ref{potential}, this is not the only inhomogeneous 
solution. However, a bounce solution towards 
negative values of $\phi$ always involves a larger action and its 
contribution can therefore be neglected. For $\Re[\theta]\neq\pi/2$
the imaginary part of the potential is finite, in general.

\begin{figure}
\centerline{\epsfxsize=2.5in\epsfbox{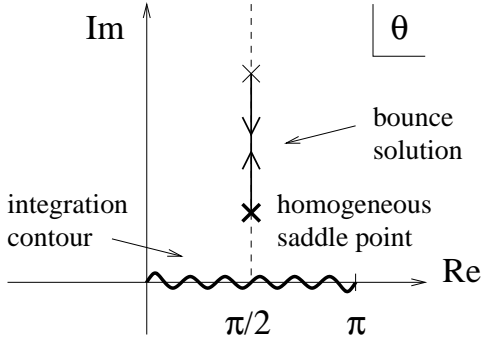}}
\caption{\label{contour} Bounce solution.}
\end{figure}

In order to obtain a finite though exponentially small sub-gap density
of states, we need to find a solution with finite action. All replica
symmetric solutions lead to a vanishing action in the limit $N\to0$.
Therefore, the solution we seek necessarily involves replica
symmetry breaking (RSB). Leaving aside the homogeneous mean-field
configuration, the configuration which incurs the lowest action is one
in which the bounce inhabits only a single replica, say, $a=1$, i.e. 
\begin{eqnarray*}
\hat{\theta}(x)=\frac\pi2+i\,{\rm diag}(\phi(x),\phi_{\sc ag},\,\dots
\,,\phi_{\sc ag})\,.
\end{eqnarray*}

In general, an analytic solution of the highly non-linear saddle-point
equation is not available. However, close to the gap edge, one can
expand $V_R(\phi)$ around the homogeneous solutions at $E_{\rm gap}$ up to
cubic order: 
\begin{eqnarray*} 
&&V_R(\phi)- V_R(\phi_{\sc ag})\nonumber\\
&\simeq&-\sqrt{6}\left(\frac{E_{\rm gap}}{\bar\Delta}\right)^{1/6}
\gamma^{1/2}(\epsilon)\,\delta\phi^2+\left(\frac{\eta 
E_{\rm gap}}{\bar\Delta}\right)^{1/3}\delta\phi^3\,,\nonumber
\end{eqnarray*} 
where $\gamma(\epsilon)=(E_{\rm gap}-\epsilon)/\bar{\Delta}$ and 
$\delta\phi=\phi-\phi_{\sc ag}$. In this limit the solution can be 
obtained analytically, namely
\begin{eqnarray*}
\delta\phi(x)=\sqrt{6}\left(\frac{\bar\Delta}{\eta^2E_{\rm
      gap}}\right)^{1/6}\gamma^{1/2}(\epsilon)\,\cosh^{-2}\left(\frac
  x{2r_{\rm drop}(\epsilon)}\right)
\,.
\end{eqnarray*}
Similarly, the action for the instanton assumes the form
\begin{eqnarray*}
S_{\rm inst.}=4\pi\nu_0\bar\Delta\xi\int_{\phi_{\sc ag}}^{\phi_{\rm max}}d\phi
\sqrt{V_R(\phi_{\sc ag})-V_R(\phi)}\,.
\end{eqnarray*}
The size of the instanton, which diverges upon approaching the gap edge, 
is given by~\cite{lo}
\begin{eqnarray}
r_{\rm drop}(\epsilon)=6^{1/4}\,\xi\left(\frac{\bar\Delta}
{E_{\rm gap}}\right)^{1/12}\gamma^{-1/4}(\epsilon)\,.
\label{r-drop}
\end{eqnarray}
The inhomogeneous solution, thus, varies on length scales much longer 
than the coherence length justifying the separation into slow and 
fast degrees of freedom in Sec.~\ref{subsec-MF}.

In higher dimensions one can assume that the bounce solution possesses
radial symmetry. Nevertheless, the problem becomes more complicated because the
saddle-point equation contains a gradient term~\cite{ZeAf00} 
\begin{eqnarray*}
\partial_{\tilde r}^2\phi+\frac{d\!-\!1}{\tilde r}\,\partial_{\tilde r}\phi+\frac12\partial_\phi
V_R(\phi)=0\,,
\end{eqnarray*} 
where ${\tilde r}=|{\bf r}|/\xi$.

However, one can still determine the parameter dependence of the
action by dimensional analysis using the Ansatz $\delta\phi=
\alpha f(|{\bf r}|/b)$. Altogether, this obtains 
\begin{eqnarray} 
S_{\rm inst.}=a_{\rm d}(\eta) \nu_0D \xi^{d-2}\gamma(\epsilon)^{(6-d)/4},
\label{Sinst}
\end{eqnarray}
where $a_{\rm d}(\eta)=c_{\rm d}\eta^{-2/3}(1-\eta^{2/3})^{-(2+d)/8}$ and 
$c_{\rm d}$ a numerical constant; $c_1=2^7\pi\sqrt{6}/5$. 

Although this completes our analysis of the profile and statistical weight
of the bounce solution, since it does not depart from the line 
$\Re[\theta]=\pi/2$, taken alone, it provides no contribution to the DoS! To 
understand why sub-gap states are associated with the bounce it is 
necessary to explore the role of fluctuations in the vicinity of the 
instanton. As emphasised in Ref.~\cite{LaSi00}, such a program turns 
out to be crucial in the present system.

%%%%%%%%%%%%%%%%%%%%%%%%%%%%%%%%%%%%%%
\subsection{Fluctuation analysis}
\label{subsec-fluct}
%%%%%%%%%%%%%%%%%%%%%%%%%%%%%%%%%%%%%%

As we have seen, the lowest energy bounce configuration involves a 
breaking of replica symmetry at the level of the mean-field. Taking
into account fluctuations in the vicinity of the bounce solution, we
will see below that there exists a zero-mode and a negative energy mode. 
The former restores global replica symmetry of the theory and, thus, 
ensures the integrity of the normalisation of the generating functional 
${\cal Z}=1$. Furthermore, the negative energy mode necessitates a 
$\pi/2$ rotation of the contour which, imparting a factor of $i$,
renders the contribution of the instanton to the DoS non-vanishing. 

To explore the influence of the fluctuations, let us return to the
quasi one-dimensional system and introduce the parameterisation
\begin{eqnarray}
Q=Re^{-W/2}\Lambda\otimes\sigma^{\sc ph}_3\otimes\sigma^{\sc cc}_3
e^{W/2}R^{-1}\,,
\label{par-fluct}
\end{eqnarray}
where $R(x)=\exp[i\sigma^{\sc ph}_1\otimes\sigma^{\sc cc}_3\theta(x)/2]$ 
is the rotation from the metallic saddle point to the inhomogeneous saddle 
point~$Q_{\sc sp}(\theta)$. The matrices~$W$ are 
subject to the symmetry condition in replica space $W_{ba}=W_{ab}^\dagger$. 
Expanding the action up to second order in the generators obtains the 
following term:
\breakon
\begin{eqnarray*}
S[W]\simeq -\frac{\pi\nu_0}8\int d{\bf r}\sum_{ab}\;{\rm tr}\,\Big[\partial 
W_{ab}\partial W_{ba}&&+\frac{1}{2}\partial\phi_a\partial\phi_b
\sigma^{\sc ph}_2 W_{ab}\sigma^{\sc ph}_2 W_{ba}+\left(F_a-\eta\cosh\phi_a
\cosh\phi_b\right)W_{ab}W_{ba}\\&&
-\eta\sinh\phi_a\sinh\phi_b\sigma^{\sc ph}_1\otimes\sigma^{\sc cc}_3 
W_{ab}\sigma^{\sc ph}_1\otimes\sigma^{\sc cc}_3 W_{ba}\Big]\,,
\end{eqnarray*}
\breakoff
\noindent
where $F_a=((\partial\phi_a)^2+V(\phi_a)-\eta\cosh2\phi_a)/2$. There are 
two types of fluctuation: 

\begin{description}

\item (a) replica-diagonal fluctuations, and

\item (b) fluctuations mixing the replicas. 

\end{description}

\noindent
Within the replica-diagonal part, the most relevant contributions are
due to fluctuations of the angle~$\phi$, i.e.~$W_{ab}=\sigma^{\sc ph}_1\otimes
\sigma^{\sc cc}_3 \varphi_a(x)\delta_{ab}$. In the $N-1$ `trivial' replicas 
($\theta=\theta_{\sc ag}$), these fluctuations are massive and, 
therefore, only lead to a weakly energy dependent prefactor. More 
important are the fluctuations which stay within the symmetry broken
replica $a=1$: 
\begin{eqnarray*}
S[\varphi_1]={1\over 2}\int dx \int dx'\,\varphi_1(x)\,\frac{\delta^2S}
{\delta\phi(x)\delta\phi(x')}\,\varphi_1(x')\,.
\end{eqnarray*}
Now this class of fluctuations has been studied extensively in the standard
literature~\cite{Coleman}. The operator $\delta^2S/\delta\phi(x)\delta\phi(x')$
has a zero-mode, $\varphi_1^{(0)}\sim\partial\phi$, due to translational 
invariance, i.e.~the action is independent of the position of the bounce. 
Furthermore, as the zero-mode is associated with a {\em bounce}
solution it has a node. 
This implies the existence of one negative energy eigen mode. To account 
for this, one has to rotate the contour away from the imaginary axis. 
This deformation of the contour provides a factor of $i$. Therefore, the 
result --- which was purely imaginary ($\Re[\theta]=\pi/2$) before --- becomes 
real and, thus, gives a finite contribution to the DoS.

Turning to the fluctuations mixing the replicas, the replica symmetry 
breaking must be accompanied by a zero-mode in replica space. Writing 
$W=W^-+W^+$, where $W^-(W^+)$ (anti-)commutes with $\sigma^{\sc ph}_1
\otimes\sigma^{\sc cc}_3$, the part of the action involving the
coupling between the replicas `1' and `$a\neq 1$' reads
\begin{eqnarray*}
S_{1a}^\pm \sim -\sum_{a\neq 1}{\rm tr}\,\left[\partial W^\pm_{1a}
\partial W^\pm_{a1}+\frac{V[\phi({\bf r})]\!-\!\widetilde V^\pm}2\, 
W^\pm_{1a}W^\pm_{a1}\right]\,,
\end{eqnarray*}
where
\begin{eqnarray*}
\widetilde V^\pm=\frac{\eta}{2}\left(\cosh2\phi_{\sc ag}+\cosh2\phi({\bf r})+
4\cosh(\phi_{\sc ag}\!\pm\!\phi({\bf r}))\right)\,.
\end{eqnarray*}

Although its presence is disguised by the choice of parameterisation,
the action involving this class of fluctuations exhibits a zero-mode. 
Indeed, its existence is made manifest by specifying the parameterisation 
$Q=UQ_{\sc sp}U^\dagger$. However, with this choice, the measure 
associated with the fluctuations becomes highly non-trivial. Nevertheless, 
it is useful for determining the 
dependence of the integration over zero-modes on the number of
replicas, $N$. Close to the gap edge, i.e.~at finite energies
$\epsilon>0$, the structure of the saddle point within the
${\sc{ph}}$- and ${\sc{cc}}$-space is completely fixed. -- The only
freedom left are rotations in replica space, $U\in{\rm U}(N)$. Thus,
dividing off the matrices which leave the saddle point invariant, the
relevant matrices $U$ belong to the coset space U($N$)/(U(1)$\times$U($N\!-\!1$)). 
Integration over the 
zero-mode gives a factor which is proportional to the volume $\sim N$ in 
the limit $N\to0$~\cite{KaMe99}. 
Or, in other words, there are $N$ saddle 
points that contribute to the integral. Using the 
parameterisation~(\ref{par-fluct}) does not change the $N$ dependence
but only the spatial structure of the zero-mode. Therefore, without 
calculating the value of the prefactor, we know that the result has 
the following form: 
\begin{eqnarray}
&&\nu(\epsilon)\sim\lim_{N\to0}\frac{1}{N}\int\! dx\,N\left[\sinh\phi(x)\!+
\!(N\!-\!1)\sinh\phi_{\sc ag}\right]\times\nonumber\\
&&\qquad\qquad\qquad\qquad\times|\chi_0(x)|^2\,e^{-S_{\rm inst.}}\nonumber\\
&&=\int\! dx\,\left[\sinh\phi(x)-\sinh\phi_{\sc ag}\right]|\chi_0(x)|^2
\,e^{-S_{\rm inst.}}\,,
\label{zero-rep}
\end{eqnarray}
where $|\chi_0(x)|^2$ describes the spatial profile of the zero-mode.

%%%%%%%%%%%%%%%%%%%%%%%%%%%%%%%%%%%%%%
\subsection{Discussion}
\label{subsec-discuss}
%%%%%%%%%%%%%%%%%%%%%%%%%%%%%%%%%%%%%%

This concludes our derivation of the sub-gap DoS: by itself the instanton
or bounce configuration provides the leading exponential dependence of 
the DoS while the fluctuations render the pre-exponential factors positive
definite. More precisely, 

\begin{description}

\item (1) the prefactor becomes real due to the negative energy eigen mode 
and the consequential deformation of the contour;

\item (2) the sub-gap DoS is non-vanishing only in the vicinity of the 
bounce as can be seen from Eq.~(\ref{zero-rep}). 

\end{description}

Altogether, taking the expression for the action in the $d$-dimensional 
system~(\ref{Sinst}), we obtain the expression for the DoS defined
by Eq.~(\ref{dos}). A further rescaling of the DoS allows an explicit
connection between the general expression and its universal 
zero-dimensional limit: to this end, let us introduce the parameter
\begin{eqnarray}
\Delta_{\rm g}=\left(\frac{2}{3}\,\bar\Delta\delta^2\right)^{1/3}
\eta^{4/9}(1\!-\!\eta^{2/3})^{1/6}\,,
\label{delta-g}
\end{eqnarray}
where $\delta=1/(\nu_0V)$ is the average level spacing of the normal system,
after which the mean-field DoS in the vicinity of the energy gap can be 
cast in the form~\cite{VaBr00}
\begin{eqnarray}
\nu(\epsilon>E_{\rm gap})=\frac1{\pi V}\sqrt{\frac{\epsilon-E_{\rm gap}}
{\Delta_{\rm g}^3}}\,.
\label{dossqrt}
\end{eqnarray}
Then, making use of Eqs.~(\ref{delta-g}) and (\ref{r-drop}), the DoS 
can be brought to the more compact form
\begin{eqnarray}
S_{\rm inst.}=\sqrt{\frac{3}{2}}\,c_{\rm d}\left(\frac{r_{\rm drop}(\epsilon)}
L\right)^d\gamma_{\rm g}(\epsilon)^{3/2}\,,
\label{s-inst}
\end{eqnarray} 
where $\gamma_{\rm g}(\epsilon)=(E_{\rm gap}-\epsilon)/\Delta_{\rm g}$
and $r_{\rm drop}$ is defined in Eq.~(\ref{r-drop}). 
Indeed, the factor $(r_{\rm drop}/L)^d$ can be further absorbed into 
$\Delta_{\rm g}$ by replacing the level spacing 
of the system, $\delta$, with the level spacing of the droplet, 
$\delta_{\rm drop}(\epsilon)=1/(\nu_0r_{\rm drop}^{\;d})$, in its definition~(\ref{delta-g}). Now, as 
noted in Ref.~\cite{VaBr00}, if the mean-field DoS exhibits a square-root
singularity of the form~(\ref{dossqrt}), fluctuations of the edge due 
to optimal fluctuations of the impurity potential are predicted to 
assume a universal form 
\begin{eqnarray}
\nu(\epsilon)\sim\exp\left[-{2\over 3}\left({E_{\rm gap}-\epsilon\over 
\Delta_{\rm g}}\right)^{3/2}\right]
\label{dosuniversal}
\end{eqnarray}
obtained by random matrix theory. Now, as we have seen, when 
$r_{\rm drop}> L$ (inevitable as $\epsilon \to E_{\rm gap}$), the system 
enters a zero-dimensional regime. In this limit, with $d=0$ the expression 
for the DoS reassuringly assumes the universal form (\ref{dosuniversal}).

In the present context, the mechanism by which the universal expression
develops at the level of the action has been elucidated in 
Ref.~\cite{LaSi01}. Specifically, in the zero-dimensional regime, the 
instanton configuration must be supplemented by a homogeneous replica 
symmetry broken solution of stationary phase which sits at the 
shallow minimum 
of the potential~$V(\theta)$, c.f.~Fig.~\ref{potential}. There the 
integration contour leaves the axis $\theta=\pi/2+i\phi$, and the minimum 
represents in fact a maximum along the perpendicular 
direction~\cite{LaSi01}. Physically, gap fluctuations in the 
zero-dimensional system correspond to sample-to-sample fluctuations 
rather than spatial variations of the gap. 

Although the saddle-point analysis as well as the size of the instanton 
agree with the result found in Ref.~\cite{lo}, the energy dependence 
of the action does not, i.e.~while we obtain the exponent $\alpha=3/2-d/4$, 
the solution obtained in Ref.~\cite{lo} is compatible with an exponent 
$\alpha_{\rm Lif}=2-d/4$. As mentioned above the exponent $3/2$ can be 
traced back to the zero-dimensional case, where the form of the action 
is universal~\cite{VaBr00}, i.e.~this exponent is a direct consequence 
of the square root behaviour of the mean-field result. 

In fact, the discrepancy of the results can be traced to the application
of a Lifshitz-type argument~\cite{Lifshitz} to the present scheme. Indeed,
although the problem bears close similarity to the Lifshitz problem of 
band tail states in semiconductors~\cite{Lifshitz,HLZL66}, the correspondence
is superficial. In particular, Lifshitz tails states at the band-edge of 
a semiconductor are typically associated with wavefunctions which vary 
smoothly on the scale of their extent. As such, an estimate of the optimal
character of the tail state distribution can be established on the level 
of the $\psi$-field action. By contrast, the sub-gap tail
states associated with gap fluctuations in the superconducting system
involve a superposition of states close to the Fermi level, where spatial
fluctuations vary rapidly oscillating at the scale of the Fermi 
wavelength --- the sub-gap states are quasi-classical in origin. It
therefore does not seem possible to develop a Lifshitz argument for
the present system. As
a further consequence, in contrast to the band-tail states, the
quasi-classical nature of the sub-gap states in the superconductor
make their properties insensitive to the nature of the impurity 
distribution.

%%%%%%%%%%%%%%%%%%%%%%
\section{Conclusions}
\label{conclusions}
%%%%%%%%%%%%%%%%%%%%%%

In summary, following the work of Larkin and Ovchinnikov~\cite{lo}, we 
have shown that, in a weakly disordered superconductor, short-scale 
fluctuations of the BCS coupling constant lead to a suppression of the
quasi-particle energy gap. At the level of mean-field, the integrity
of the gap edge is maintained. However, optimal fluctuations of the
impurity potential induce a narrow band of states, localised at the 
scale of the coherence length, which extend below the mean-field gap 
edge. Within the framework of the statistical field theory developed 
here, these states appear as replica symmetry broken instanton 
configurations of the mean-field equations --- the global symmetry of
the theory being restored by a zero-mode in the replica space. To
exponential accuracy, we have obtained the spectrum of gap fluctuations. 
The generality of these results has been emphasised. Specifically, 
in the $d$-dimensional system, once normalised by the mean-field DoS 
at the gap edge, we have shown that the spectrum of tails states 
depends only on the dimensionless parameter $\eta$ and, in particular, 
is independent of the nature of the disorder potential. Moreover, in 
the zero-dimensional system, the spectrum of gap fluctuations is truly 
universal and coincides with that obtained by Vavilov 
{\em et al.}~\cite{VaBr00} in the study of gap fluctuations in the SN 
system.

Finally, it is interesting to note that the analysis in this work has
a number of relatives in the recent literature. As we have emphasised,
at the level of the soft mode action, the theory of gap fluctuations 
mirrors that obtained in the Abrikosov-Gor'kov theory of a
superconductor with magnetic impurities~\cite{LaSi00,LaSi01} and,
later, that encountered in the description of sub-gap states in the
hybrid SN system~\cite{feigelmann00}. Furthermore, various results 
non-perturbative in the (inverse) dimensionless conductance and
involving replica (or super-)symmetry breaking have been reported in
the literature~\cite{em,km,fe,Kamenev}. Of these investigations, it is
particularly interesting to contrast the present scheme with the
prediction of `anomalously localised states' in the weakly disordered 
normal conductor. 

By exploiting instanton configurations of the 
non-linear $\sigma$-model action, Khmel'nitskii and 
Muzykantskii~\cite{km} proposed that the long-time current relaxation 
in a disordered wire was dominated by rare localised states which
coexist in a background of extended states (see also, Ref.~\cite{fe}). 
These states, which are ascribed to optimal fluctuations of the random
potential, are penalised by a statistical weight which depends
exponentially on the dimensionless conductance. This scaling mirrors
that found in the present system. However, crucially, scaling in the
superconducting system involves an energy dependence which allows the
exponent to become small as one approaches the energy gap. 

In hindsight, it is easy to understand why optimal fluctuations can
more readily induce localised states in the superconducting system. In
the normal disordered system, as pointed out by Mott, hybridisation 
makes the coexistence of localised states in a background of extended
states difficult to sustain. However, in the superconducting system,
fluctuations of the order parameter provide a natural mechanism by
which quasi-particle states can localise in regions where the order 
parameter is suppressed. 

%%%%%%%%%%%%%%%%%%%%%%%%%%%%%%%%%%%%%%%%%%%%%%%%%%%%%%%%%%%%%%%%%%%%%%%%%%%%%%%
%%%%%%%%%%%%%%%%%%%%%%%%%%%%%%%%%%%%%%%%%%%%%%%%%%%%%%%%%%%%%%%%%%%%%%%%%%%%%%%

{\sc Acknowledgements:} We are grateful to Dima Khmel'nitskii and Austen
Lamacraft for useful discussions.

%%%%%%%%%%%%%%%%%%%%%%%%%%%%%%%%%%%%%%%%%%%%%%%%%%%%%%%%%%%%%%%%%%%%%%%%%%%%%%%
%%%%%%%%%%%%%%%%%%%%%%%%%%%%%%%%%%%%%%%%%%%%%%%%%%%%%%%%%%%%%%%%%%%%%%%%%%%%%%%

\end{multicols}
\end{document}